\documentclass[reprint,groupedaddress]{revtex4-1}
\usepackage{amsmath}
\usepackage{graphicx}
\usepackage{epstopdf} 
\usepackage{holtpolt}
\usepackage{amstext,bbold}

\usepackage{enumitem}
\setitemize{label=\textbullet, leftmargin=*, nolistsep}

\begin{document}

\title{A Proposed Experimental Search for Chameleons using Asymmetric Parallel Plates}

 \author{Clare Burrage}
 \email{Clare.Burrage@nottingham.ac.uk}
 \affiliation{School of Physics and Astronomy, University of Nottingham, Nottingham, NG7 2RD, United Kingdom}
 \author{Edmund J. Copeland}
 \email{ed.copeland@nottingham.ac.uk}
 \affiliation{School of Physics and Astronomy, University of Nottingham, Nottingham, NG7 2RD, United Kingdom}
 \author{James A. Stevenson}
 \email{james.stevenson@nottingham.ac.uk}
 \affiliation{School of Physics and Astronomy, University of Nottingham, Nottingham, NG7 2RD, United Kingdom}

\date{\today}

\begin{abstract}
Light scalar fields coupled to matter are a common consequence of theories of dark energy and attempts to solve the cosmological constant problem.  The chameleon screening mechanism is commonly invoked in order to suppress the fifth forces mediated by these scalars, sufficiently to avoid current experimental constraints, without fine tuning. The force is suppressed  dynamically by allowing the mass of the scalar to vary with the local density. Recently it has been shown that near future cold atoms experiments using atom-interferometry have the ability to access a large proportion of the chameleon parameter space. In this work we demonstrate how experiments utilising asymmetric parallel plates can push deeper into the remaining parameter space available to the chameleon.

\end{abstract}

\maketitle

The current cosmological model is unable to describe  what is driving the accelerated expansion of our universe, without invoking the extreme fine tuning associated with the cosmological constant. Scalar fields are often introduced either  directly to  explain the observed accelerated expansion of the universe, or indirectly as a consequence of attempts to solve the cosmological constant problem \cite{Copeland:2006wr,Clifton:2011jh,Joyce:2014kja}.  
Generically, adding a light scalar field
% capable of generating cosmological expansion 
would induce a fifth-force in conflict with a host of stringent experimental constraints \cite{Adelberger:2003zx}. This can be alleviated  through the presence of a screening mechanism which dynamically suppresses the fifth force in   experiments  through self interactions of the scalar field \cite{Joyce:2014kja}. The most widely studied model of this type is the chameleon \cite{Khoury:2003aq,Brax:2004qh}, which inherits a density dependent effective potential via a conformal coupling to standard model matter fields making the  field mass increase with the density of the environment, thus the chameleon mediated force  becomes increasingly short range. 
%As a result, fifth-force interactions can be screened in the relatively high density of our solar system but survive the lower densities on cosmological scales. 

The fact that the chameleon theory has been designed to suppress the associated fifth-force in dense environments, means that it could in principle be detected in a suitably designed laboratory experiment in high vacuum.  Recent years have seen a wealth of experiments formulated for or appropriated to the detection of dark energy, reaching far into previously unconstrained regions of the theory parameter space utilising atom-interferometry \cite{Burrage:2014oza,Burrage:2015lya,Hamilton:2015zga}, cold neutron experiments \cite{Brax:2013cfa,Ivanov:2012cb,Lemmel:2015kwa,Li:2016tux} and searches for Casimir-like forces \cite{Almasi:2015zpa,Brax:2014zta}. The remaining parameter space includes the chameleon theories that are most weakly coupled to matter, making them intrinsically harder to detect. 
%This is because the more strongly coupled a chameleon is, the more efficient its screening mechanism \cite{Mota:2006fz}. 

The target of this work is to understand whether the non-linearities of the chameleon field can be exploited in order to allow experiments to probe the  remaining parameter space.  In this letter we propose a  parallel plate set-up where the two plates have different densities but the same total mass. This results in an asymmetric field profile for the chameleon field, but a symmetric field profile for the gravitational force.  In this regime of the parameter space, the chameleon force depends intrinsically on the plate density, becoming independent of the total mass of the plate.  This asymmetry allows us to enhance the chameleon presence in an atomic interferometry experiment with the additional advantage of removing the gravitational background. 

\section*{The Chameleon Theory}\label{The_Chameleon_Theory}

The chameleon scalar field is  described by the following action:
\begin{equation}
\begin{split}
\mathcal{S} = \int d^4x \sqrt{-g} \left\{\frac{M_{Pl}^2}{2}R - \frac{1}{2}\partial_\mu\phi\partial^\mu\phi - V(\phi)\right\} \\ + \int d^4x \>\mathcal{L}_m \big(\psi_{m}, \Omega^{2}(\phi) g_{\mu\nu}\big) \;,
\end{split}
\end{equation}
This  assumes a universal, minimal coupling between the conformally rescaled metric $\tilde{g}_{\mu\nu} = \Omega^{2}(\phi)g_{\mu\nu}$ and each matter species $\psi_m$. The conformal factor $\Omega(\phi)$ is chosen to be a monotonically increasing function of the chameleon field $\phi$, which we approximate by $\Omega(\phi) = 1+ \phi/M$. It can be checked that in all situations we consider   $\phi \ll M$. The energy scale $M$  controls the coupling to matter, and  $M=M_P$ indicates a gravitational strength force.   The chameleon relies on a potential containing non-trivial self interactions, for concreteness in this work we will take $V(\phi) = \frac{\Lambda^5}{\phi}$ which captures all of the interesting chameleon phenomenology.
% for example any power law form  $V(\phi)= \Lambda^{4+n} \phi^{-n}$ excepting $n=-1,-2$.   
The energy scale  $\Lambda$  controls the self interactions of the scalar.  The scale $M$ is constrained, by precision measurements of atomic structure and searches for fifth forces \cite{Mota:2006fz,Brax:2010gp,Upadhye:2012qu}, to lie in the range $10^{4}\mbox{ GeV} \leq M \leq M_P \sim 10^{18} \mbox{ GeV}$. Through the connection to dark energy and the increasing expansion rate of the universe, $\Lambda$ is expected to be of order $1\mbox{\,meV}$ \cite{Ade:2013zuv}, while Casimir force measurements indicate  $\Lambda <100 \mbox{\,meV}$ \cite{Mota:2006fz,Gannouji:2010fc,Upadhye:2012qu}. Given these constraints we take $10^{-2} \mbox{\,meV}<\Lambda<10^{+2} \mbox{\,meV}$ as our  range  of interesting values. Additional constraints on the parameter space from atom interferometry \cite{Hamilton:2015zga} are presented alongside our key results in Figures \ref{phase_difference} and \ref{phase_difference_largel}.

When matter can be described as a static, non-relativistic perfect fluid, the motion of the  chameleon is governed by an effective potential 
\begin{equation}
V_{\rm{eff}} = \frac{\Lambda^5}{\phi} + \frac{\rho\phi}{M}
\end{equation}
where $\rho$ is the local energy density of matter. 
%Substituting  the specific forms of  $V(\phi)$ and  $\Omega(\phi)$ described above, we  find that 
The minimum of the effective potential,  $\phi_{\rm min}(\rho)$, and the mass of fluctuations around this minimum, $m_{\phi}(\rho)$ are given by
\begin{equation}
\phi_{\rm{min}}(\rho) = \Big(\frac{\Lambda^5 M}{\rho}\Big)^{1/2}, \>\>\>\>\> m_{\phi}^2(\rho) = 2\Big(\frac{\rho^3}{\Lambda^5M^3}\Big)^{1/2}
\label{eq:min}
\end{equation}
The force law associated with this coupling is
$F_{\phi} = -\frac{1}{M}\nabla\phi$.
Equation \eqref{eq:min} shows how the mass of the chameleon field increases with the density, allowing the field to evade fifth force constraints \cite{Khoury:2003aq} under certain conditions.  
%In this work we present a means to disentangle the chameleon and gravitational interactions in an attempt to isolate the chameleon force. Our approach focuses on exploiting the chameleon's  response to the density of the source, allowing us to devise a scenario where the gravitational background vanishes, we will find that this method  unlocks previously inaccessible portions of the chameleon parameter space.

\section*{The Parallel Plate Set-up}

\begin{figure}[]
\centering
\vspace{0.75 cm}
\centerline{\scalebox{0.48}{\includegraphics[trim={1cm 18.5cm 0cm 2cm}]{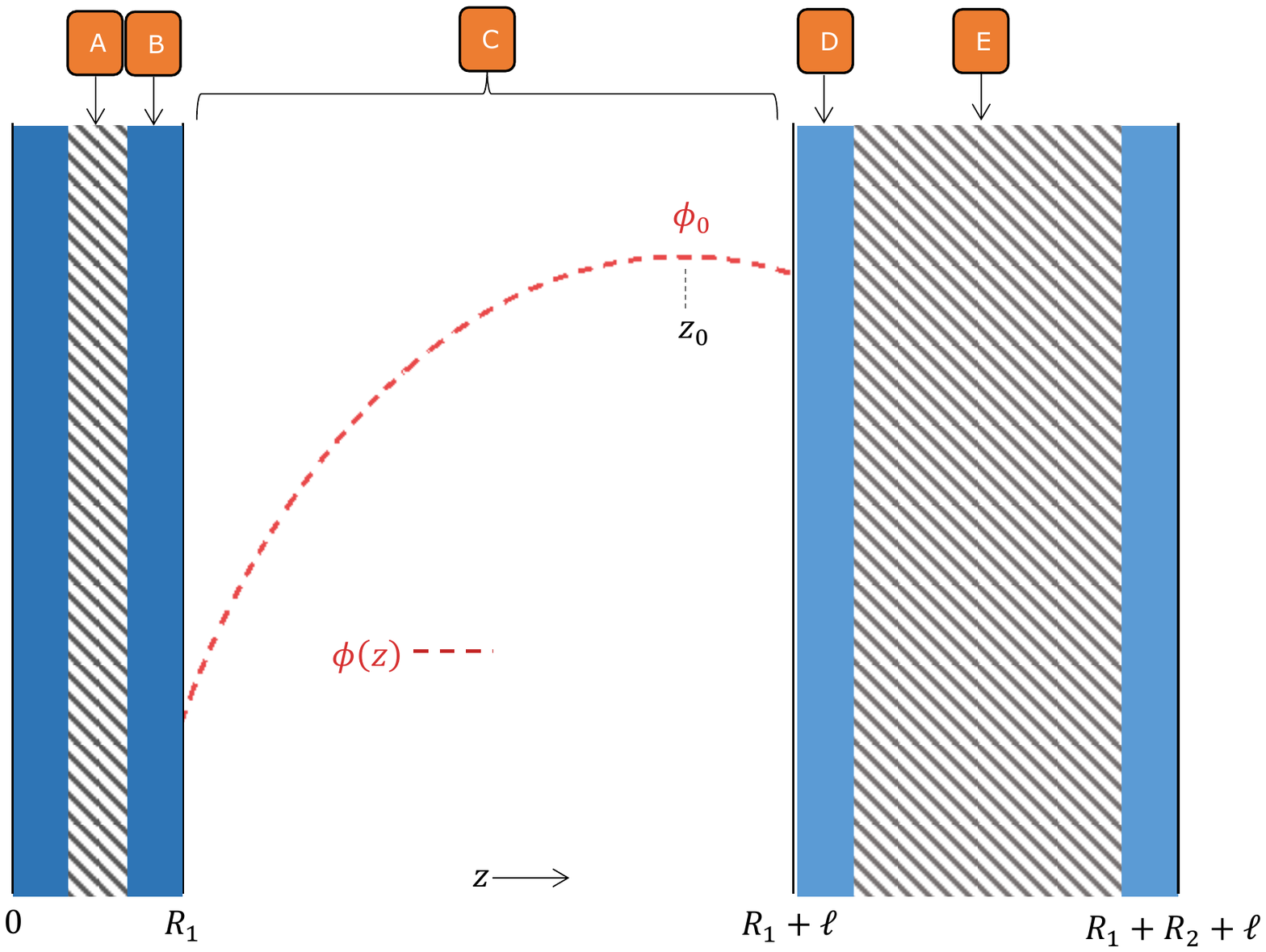}}}
\vspace{1.3 cm}
\caption{Parallel plate configuration with contrasting density values, keeping the density multiplied by the width of the plate the same for both plates.  The plate separation is denoted $\ell$ with the left and right plate thickness's as $R_1$ and $R_2$, with corresponding plate densities $\rho_L$ and $\rho_R$  respectively. The regions shaded with lines indicate where the field has reached the minimum of its effective potential inside the plate.  The solid shaded regions  are the shells in which the field rolls away from the minimum of the effective potential.  }\label{Plate_Setup}
\end{figure}

We consider the chameleon field between two parallel plates, and allow the plates to have differing densities.  We make the approximation that the plates are infinite in extent, and that the space in between the plates has almost zero density. The gravitational potential between the plates is therefore approximately constant, and so the gravitational force vanishes.  
In contrast, we will find that the chameleon has a non trivial field profile between the plates, as can be seen in Figure \ref{Plate_Setup}, and that this form is a function of the density of the two plates. 
Through out this work, if the densities of the plates differ we will assume without loss of generality that the left most plate has the higher density. We consider two possibilities, the first that the field reaches a maximum between the plates, and the second that the field is monotonically increasing in the space between the plates.

\subsection*{Chameleon field reaches a maximum between plates}
Analogous to the thin shell effect that occurs for the chameleon field profile around a sphere, if the plates are sufficiently wide they can be divided into core and shell regions. Represented by bands A and E in Figure \ref{Plate_Setup}, the core defines the space where the field $\phi\simeq \phi_{\rm min}$ of Equation (\ref{eq:min}). Inside the plate the field only varies in the shell (bands B and D), increasing from the value that minimises the effective potential in the plate. In doing so it traverses the part of the effective potential that is dominated by the density contribution. In the low density region between the two plates (band C)  the chameleon field will evolve on the part of the effective potential controlled by the self-interaction terms of the bare potential. 
Importantly, as long as core regions exist for both plates, the chameleon field between the plates becomes effectively independent of the behaviour of the chameleon outside the plates, any chameleon perturbation attempting to propagate through the wall becomes exponentially suppressed if the Compton wavelength of the chameleon inside the plates is smaller than the plate thickness.
We denote the thickness of the shell in the left and right plates as $\delta_L$ and $\delta_R$ respectively and impose that the field and its first derivative are continuous at the surface of the plates and at the interface between the shell and core regions. 
The problem therefore  reduces to the following set of equations:
\setitemize{label=\textbullet, leftmargin=*, nolistsep}
\label{Solutions_Set_General}
\begin{itemize}
\item[]{\underline{$R_1 - \delta_L < z < R_1$ }}
\begin{equation}\label{eq:shell_left} \phi(z) = \frac{\rho_L}{2M}\left(z - (R_1 - \delta_L)\right)^2 + \phi_{L}
\end{equation}  
\item[]{\underline{$R_1  < z < R_1 + \ell$ }}
\begin{multline}\label{eq:implicit}
\rm{arccos}\sqrt{\frac{\phi(z)}{\phi_0}} + \sqrt{\frac{\phi(z)}{\phi_0}}\left(1 - \frac{\phi(z)}{\phi_0}\right)^{\frac{1}{2}} \\ = \mp \sqrt{\frac{2\Lambda^5}{\phi_0^3}}\left(z - z_0\right)
\end{multline}
\item[]{\underline{$R_1 + \ell  < z < R_1 + \ell + \delta_R $ }}
\begin{equation}\label{eq:shell_right}
\phi(z) = \frac{\rho_R}{2M}\left(z - \left(R_1 + l + \delta_R\right)\right)^2 + \phi_{R}
\end{equation}
 \end{itemize}
the constant parameter $z_0$ corresponds to the position  at which the chameleon field reaches its maximum value, denoted $\phi_0$, see Figure \ref{Plate_Setup}.  We have written  $\phi_L = \phi_{\rm min}(\rho_L)$ and below we introduce $m_L= m_{\phi}(\rho_L)$. The equivalent definitions hold for the right hand plate.

Imposing boundary conditions at the surfaces of the two plates we find four equations for the four unknowns $\phi_0$, $z_0$, $\delta_L$ and $\delta_R$:
\begin{equation}\label{eq:polynomial}
\frac{\phi_i}{\phi_0}= \frac{1-\left(\frac{m_i\delta_i}{2}\right)^2-\left(\frac{m_i\delta_i}{2}\right)^4}{1+\left(\frac{m_i\delta_i}{2}\right)^2}
\end{equation}
where $i\in(L,R)$ and 
\begin{multline}\label{eq:implicit_new}
\arccos \sqrt{\frac{1}{1+\frac{\phi_0}{\phi_i}\left(\frac{m_i\delta_i}{2}\right)^2}}+\frac{\sqrt{\frac{\phi_0}{\phi_i}}\frac{m_i\delta_i}{2}}{1+\frac{\phi_0}{\phi_i}\left(\frac{m_i\delta_i}{2}\right)^2}\\= \left(\frac{\phi_i}{\phi_0}\right)^{3/2}\left\{\begin{array}{lc}
m_L(z_0-R)& \mbox{ if } i=L\\
m_R(R+l-z_0) & \mbox{ if } i=R
\end{array}\right.
\end{multline}
If the field reaches a maximum value between the plates then there is a smallest such maximum set by  $\phi_0 \geq \phi_R$. The system above can be solved by first considering the series expansion of \eqref{eq:implicit_new} for small $\phi_L / \phi_0$ which can be used to give a closed form solution for the shell thickness $\delta_L$
\begin{equation}
\left(\frac{m_L \delta_L}{2}\right)^2 = \frac{1}{2}(\sqrt{5} -1)
\label{eq:deltaL}
\end{equation}
This assumption is found to remain consistent across the entire chameleon parameter space provided that the following density inequality is satisfied
\begin{equation}
\sqrt{\frac{\rho_L}{\rho_R}} \gtrsim \frac{\sqrt{5} + 1}{2}
\end{equation}
Equation \eqref{eq:deltaL} can then be used to recover an explicit equation for $z_0$
\begin{equation}
z_0 = R_1 + \frac{2}{3m_L}\left\{\frac{3\pi}{4}\left(\frac{\phi_0}{\phi_L}\right)^{3/2} - \left(\frac{2}{\sqrt{5} - 1}\right)^{3/2}\right\}
\label{eq:z0}
\end{equation}
%We can allow the density of the RH plate to vary freely, as long as a minimum density ratio is met such that the approximation leading to the above is satisfied.   
If $\phi_R\ll \phi_0$ then there exists a similar set of equations to (\ref{eq:deltaL})-(\ref{eq:z0}) relating $\delta_R$, $z_0$ and $\phi_0$,  and the field profile between the plates is approximately symmetric.  The other option is to take $\phi_R \approx \phi_0$. This will be the case when the field configuration is extremely asymmetric and  $z_0$ lies at the surface of  the right hand  plate. For the more general case, equations \eqref{eq:polynomial} and \eqref{eq:implicit_new} should be solved numerically. 
The key observation here, is that for any given value of the chameleon self interaction scale $\Lambda$ the degree of asymmetry in the chameleon field profile (following from the value of $z_0$ being offset from the central position of $R_1 + \ell/2$) between the plates increases as the value of the chameleon energy scale $M$ is increased. This indicates that asymmetry effects may allow us to probe the weak coupling regime, where the associated fifth force will be more difficult to detect. 
%Fixing the experimental parameters,  $\ell, \rho_L$ and $\rho_R$, the regions of parameter space which lead to an extreme asymmetry can be seen in the red region of Figure \ref{Parameter_Space_Collective}. 
%The remaining white regions of parameter space cannot be treated  analytically, but will interpolate between these two regimes. \\
 
\subsection*{No maximum for the chameleon field}
The maximum amount of asymmetry possible if the chameleon field reaches a maximum value between the plates occurs when $z_0= R_1 +\ell$. In this section we consider what happens if the field has insufficient space to relax into a maxima within the vacuum region.  In that case the field value continues to rise within the right hand plate, and its evolution is further described by equation (\ref{eq:implicit}). It follows that equation \eqref{eq:shell_right} can be omitted from the calculation (with the unknown $\delta_R$ being no longer relevant to the problem) and the upper bound of Equation \eqref{eq:implicit} can be set to $z_0$, which now lies inside the plate. If the plate is sufficiently wide the field continues to increase until it reaches the value that minimizes the effective potential in the right hand plate, specifying the value of the field maxima to be $\phi_0 = \phi_R$. The system then reduces to solving equations \eqref{eq:polynomial} and \eqref{eq:implicit_new} across the single boundary corresponding to the surface of the left-hand plate.

\section*{Atom Interferometry}
The asymmetric chameleon field profile  will produce a force on any particles travelling between the plates which could be detected in a sufficiently sensitive experiment. One could perform a classical deflection experiment, firing a beam of particles between the plates, and measuring the deviation from straight line motion in order to constrain the chameleon.  Recently it has been shown that experiments using atom interferometry are particularly sensitive probes of the chameleon field \cite{Burrage:2014oza,Hamilton:2015zga,Burrage:2015lya}. This is due to the fact that in a laboratory vacuum atoms are not screened from the chameleon field over a broad region of the chameleon parameter space, in addition the nature of the experiment means that it can be sensitive to extremely small forces acting on the individual atoms. This motivates combining our asymmetric plate scenario with an atom interferometry measurement.

\begin{figure}[]
\centering\vspace{0.5 cm}
\centerline{\scalebox{0.65}{\includegraphics[trim={1cm 0.5cm 0 0.5cm}]{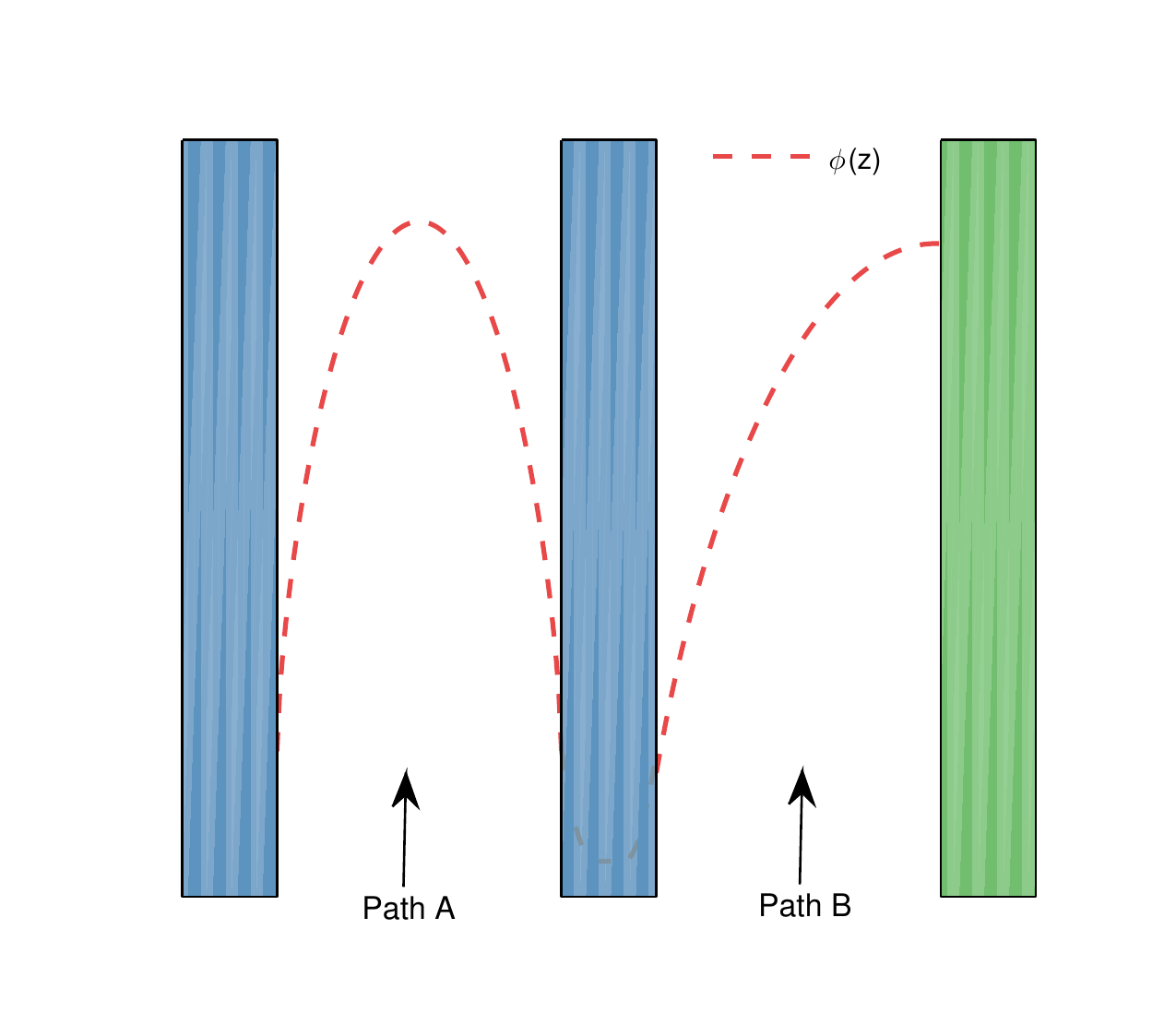}}}
\caption{The three plate configuration proposed for use in an atom interferometry experiment.  The right most (green) plate is less dense than the left and central (blue) plates, giving rise to the chameleon field profile indicated by the red dashed line.}\label{int_setup}
\end{figure}

Atom interferometry  divides the wave function of an atom  into  a superposition of two states, which  traverse spatially separated  paths before later being recombined. The phase of each state becomes a function of the path that has been traversed, and when the two states are combined the differing phases of the two possible states of the wave-function result in an interference pattern.  The probability of measuring the atom to be in one of the two states therefore becomes a function of the forces that have acted on the atom during the evolution. 
We now  demonstrate the capability of an experiment adopting an asymmetric plate configuration.
Our results are based upon a set-up where one of the two paths explored by the wave-function traverses an asymmetric chameleon profile whilst the other path passes between a symmetric plate configuration, as shown in Figure \ref{int_setup}.  To simplify the calculation we assume that incident atomic beams are taken to enter exactly halfway between neighbouring plates. This allows for the symmetric path to be used as a reference, as particles avoid both gravitational and chameleon forces. 

The experiment measures the phase difference accumulated along the two paths which has two parts.  The first is proportional to the difference in the classical action evaluated along each path \cite{feynman,storey:jpa-00248106}, and the second is imprinted by the interactions with the laser beams used to manipulate the atoms. Basically, the atoms pick up a phase proportional to $k_{\rm{eff}}$ for each interaction where $k_{\rm{eff}}$ is the effective wave-number associated with the hyperfine splitting transition of the atom.  Due to the implicit form of equation (\ref{eq:implicit}) describing the chameleon field between the plates, one needs to resort to numerical methods to evaluate these phases. However it is possible to obtain an analytic estimate for this integral, if the deviation of the particle path from a straight line is small. In that case we can consider the expansion of the scalar field to linear order around its central value. Performing this calculation, the  contribution to the atomic phase difference from the classical action, denoted $\Delta^A$, can be approximated by
\begin{equation}\label{eq:Phase_Diff_A}
\Delta^A \approx \left\{\frac{1}{3}a^2 T^2 + \frac{1}{M}\left(\phi_{B} - \phi_{A}\right)\right\}T
\end{equation}
where $a$ is the constant chameleon acceleration experienced throughout the asymmetric route, $T$ is the net exposure time and the quantities $\phi_{B}$ and $\phi_{A}$ correspond to the central field values of the right hand and left hand paths respectively.  Assuming that the atoms are kicked by the lasers just before they enter the plates, to put them on the correct path, and just after exiting the plates, in order to recombine them, the net contribution from these interactions $\Delta^{P}$ is given by:
\begin{equation}\label{eq:Phase_Diff_P}
\Delta^{P} = k_{\rm{eff}}aT^2
\end{equation}
Combining \eqref{eq:Phase_Diff_A} and \eqref{eq:Phase_Diff_P}, the net phase difference can be seen to consist of two competing factors: one that depends on the  acceleration (proportional to the gradient of the chameleon field)  and one that depends on the scalar potential.  As illustrated in Figure \ref{int_setup}, the central acceleration can be increased by increasing the density asymmetry but this is at the expense of decreasing the central field value. Conversely, the phase difference due to the scalar potential terms can be increased by allowing the separation between the two sets of plates to vary.  In moving to a larger plate separation for path B, the field is able to reach a higher central value.     

%Either, the acceleration can be maximised at the expense of shifting the central field value, or the contribution due to the scalar potential can be maximised at the cost of acceleration.  These two possibilities translate into two experimental configurations. The former, relies on an asymmetric field profile which favours a smaller plate separation. Whereas the latter cares more about an asymmetry in the plate separations which feeds directly into a contrast in the central field values. 
 
\begin{figure*}[]
\centering\vspace{0.5 cm}
\begin{minipage}{0.48\textwidth}

\includegraphics[width = \linewidth]{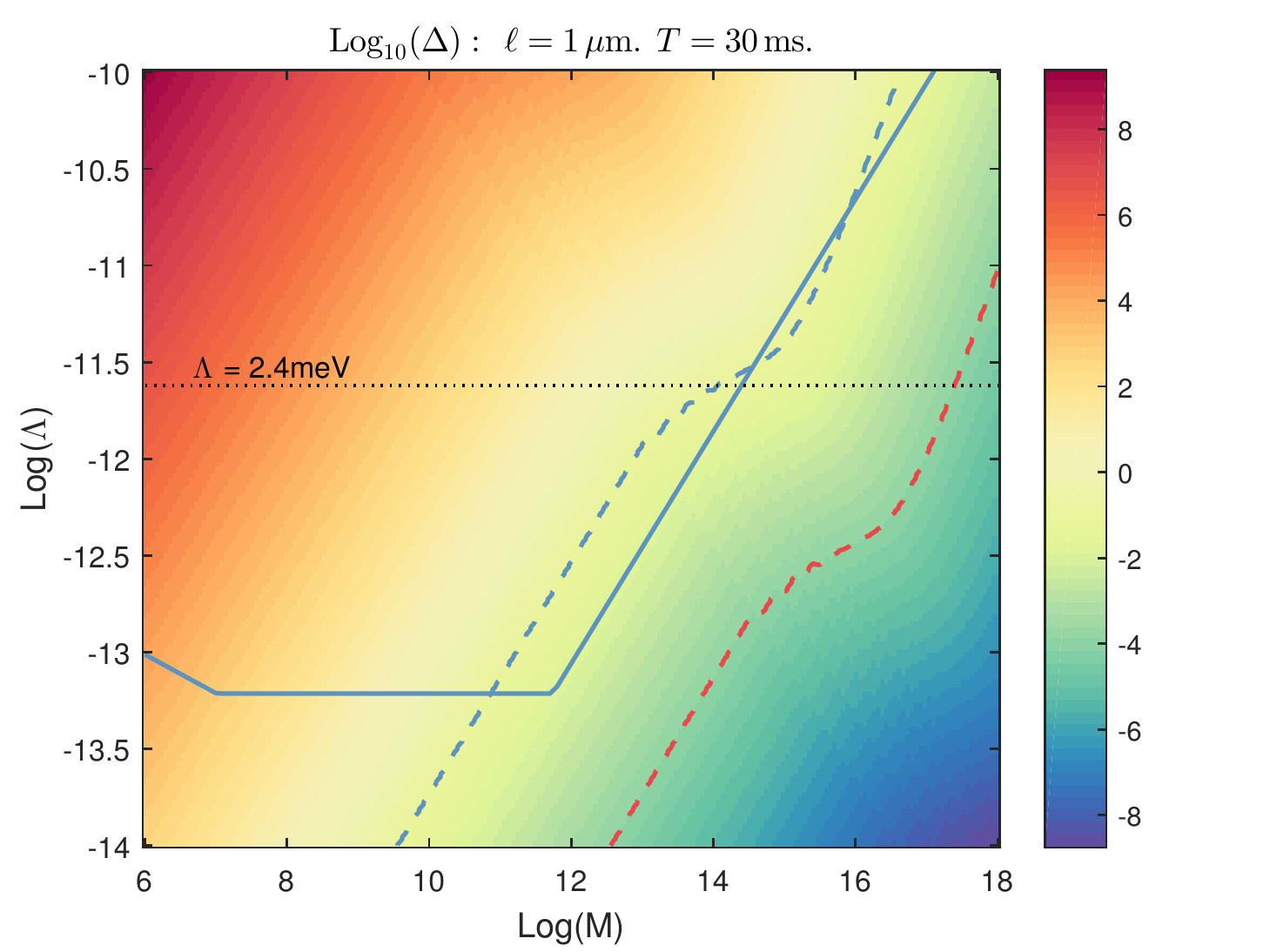}
\end{minipage}
\begin{minipage}{0.48\textwidth}
\includegraphics[width = \linewidth]{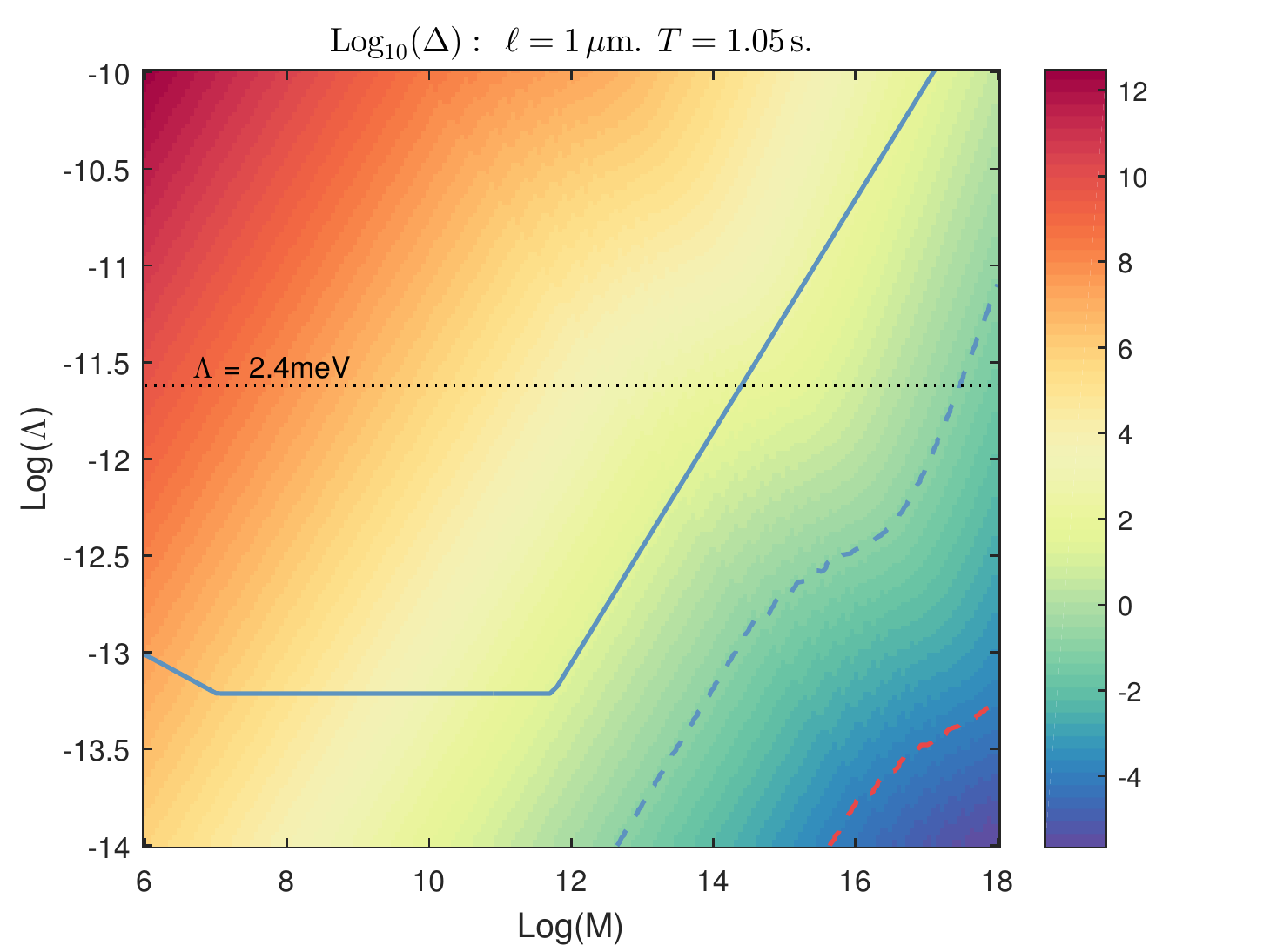}
\end{minipage}
\caption{ The phase difference $\Delta = \Delta^A + \Delta^P$ of the wave-function accumulated along the two paths indicated in Figure \ref{int_setup}, for plates separated by $\ell = 10^{-6}\,\mbox{m}$. Values for the interferometry experiment are chosen to be those of the configuration described in \cite{Burrage:2014oza}.  We assume that the experiment is placed centrally in a $10\mbox{\,cm}$-radius vacuum chamber containing $10^{-10}\mbox{\,Torr}$ of hydrogen. Moving left to right corresponds to an increase in the exposure time $T$ from $30$\,ms to 1.05\,s. Regions above the blue dashed line could be excluded by an experiment sensitive to accelerations down to $10^{-6}\,\rm{g}$, at which level systematic errors can be neglected.  Moving to an experiment sensitive to $10^{-9}\,\rm{g}$  would push the exclusion limits down to the red dashed line. The region above the solid line has been excluded in \cite{Hamilton:2015zga}.}\label{phase_difference}
\end{figure*}
\begin{figure*}[]
\centering\vspace{0.5 cm}
\begin{minipage}{0.48\textwidth}
\includegraphics[width = \linewidth]{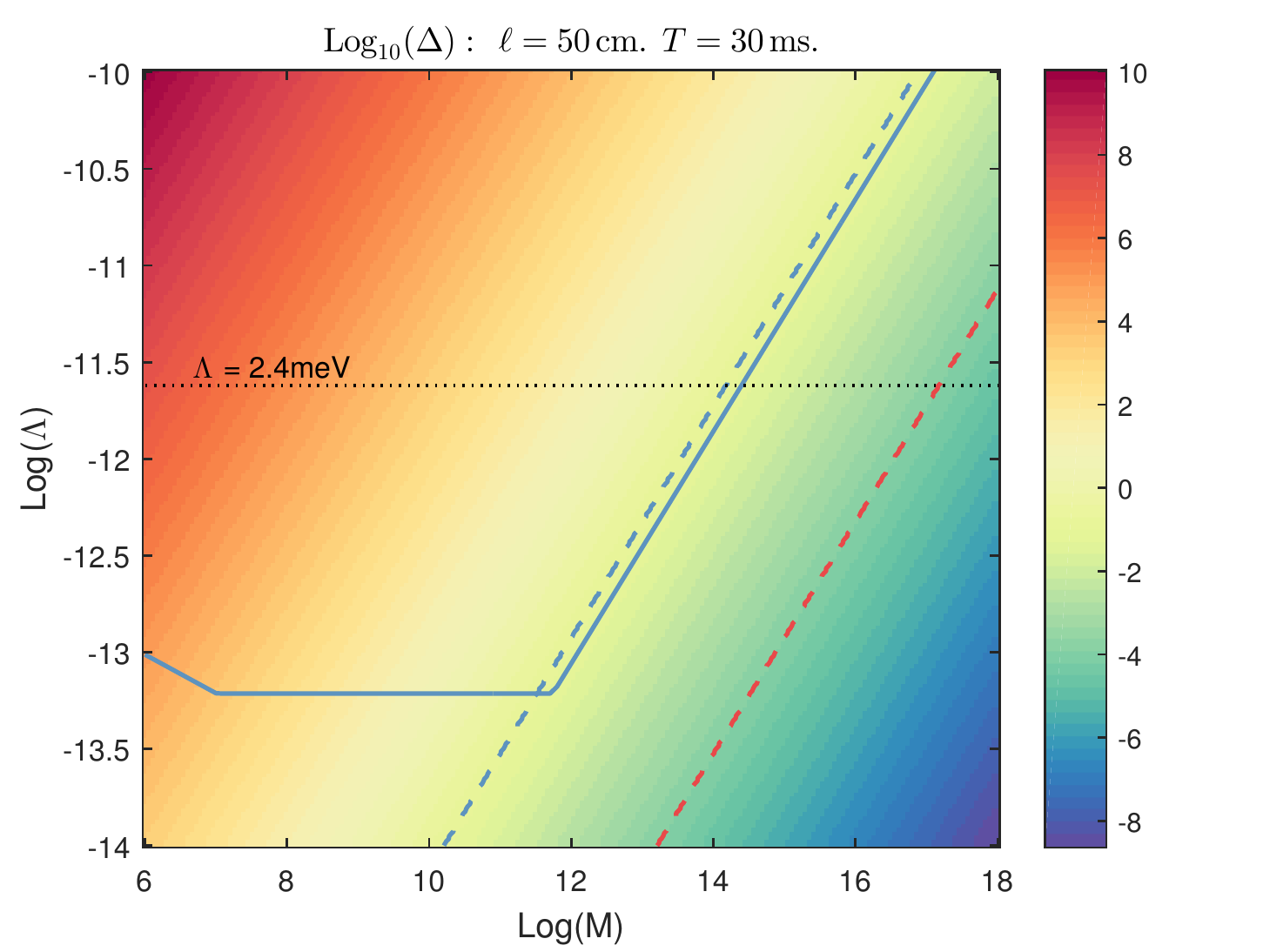}
\end{minipage}
\begin{minipage}{0.48\textwidth}
\includegraphics[width = \linewidth]{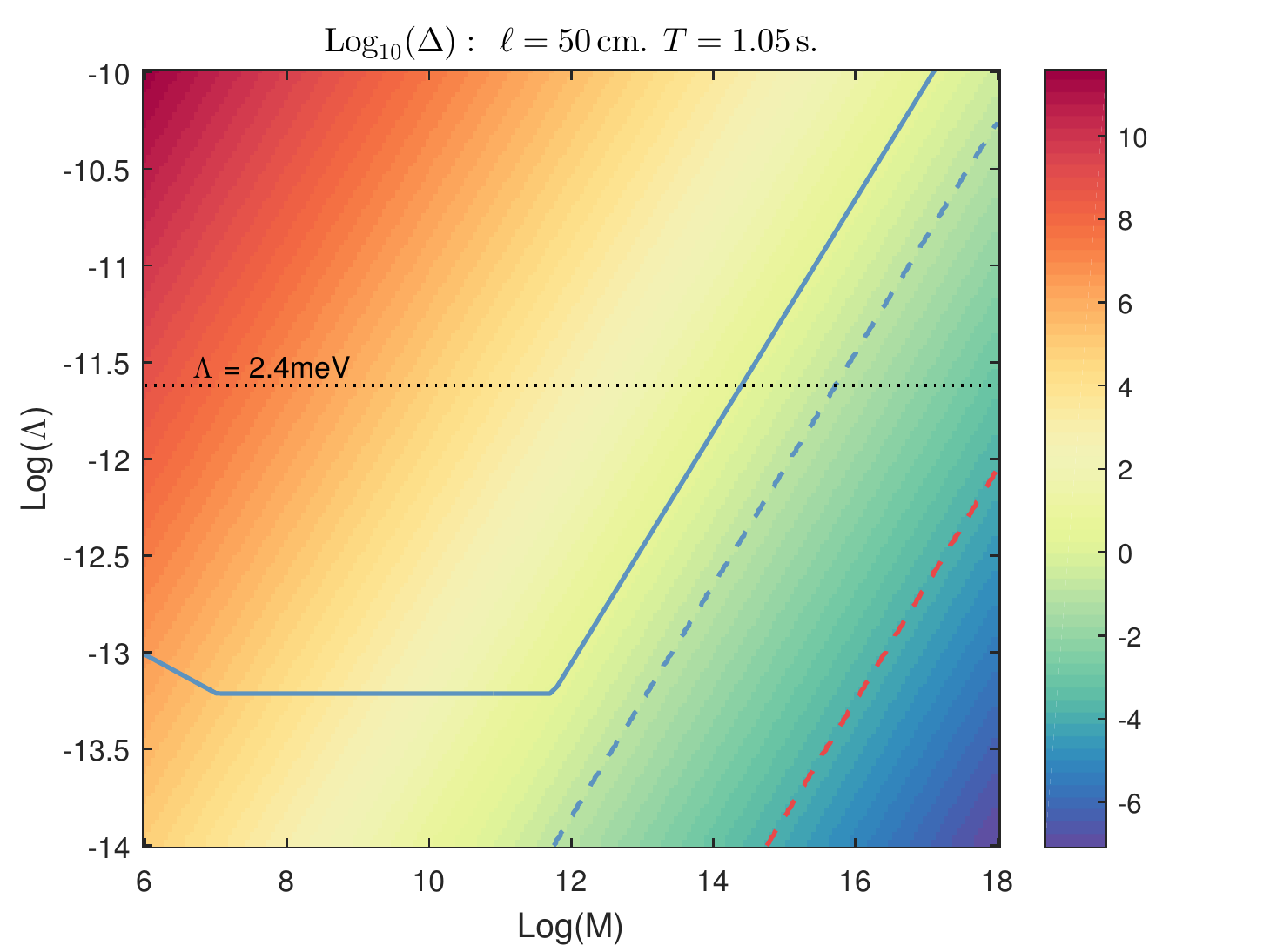}
\end{minipage}
\caption{ As in Figure \ref{phase_difference} except for the more optimistic plate separation of $\ell = 0.5\,\mbox{m}$. This would require a beam coherence length of approximately 25\,\rm{cm} which falls within the upper bound proposed in \cite{Kovachy:2015}. Current atom interferometry constraints fall on the solid blue line, which would extend to cover the region above the dashed blue line by utilising an asymmetric parallel plate configuration. These limits would shift further to the dashed red line for detectors reaching a precision of $10^{-9}\,\rm{g}$.}
\label{phase_difference_largel}
\end{figure*}
Figures \ref{phase_difference} and \ref{phase_difference_largel} demonstrate the power of this experiment for two choices of plate separation, and time of flight of the atoms between the plates. %distinct examples?
Experimental parameters for the vacuum chamber are chosen to correspond to those used in \cite{Burrage:2014oza}. Further, we assume that the atomic beams are composed of Caesium atoms. Moving from left to right in both of these figures indicates an increase in exposure time from $T = 30$\,ms, to the more optimistic $T = 1.05$\,s proposed in \cite{Kovachy:2015}. 
%For each plate  separation we find  a sizeable portion of the parameter space where the chameleon force is  sufficiently high to cause the atoms to collide with one of the plates,
%and the extent of this region grows as the plate separation is reduced. We expect that this would be observable in any experimental scenario, however for such solutions we lose control of the evolution of the atom once it has collided with the plate, and therefore we don't compute the interferometry phase for such trajectories. 
A first experiment is anticipated to be able to measure accelerations down to $10^{-6}$ g, where g is the acceleration due to free-fall at the surface of the Earth. If systematics can be controlled it is possible that a sensitivity of $10^{-9}\,\rm{g}$  could be achieved.  

\section*{Summary}
We have described an asymmetric  parallel plate set up which could be used to search for chameleon dark energy.  The asymmetry of the chameleon field profile increases for the most weakly coupled chameleons,  allowing us to overcome the difficulties of detecting the force mediated by such weakly coupled fields.   It can be seen from Figure \ref{phase_difference} that this covers most of the remaining chameleon parameter space, and in particular that the asymmetry allows us to push further into the weakly coupled (high M) region of the parameter space.   Additionally, such an experiment would improve constraints at small $M$ and $\Lambda$ which are also hard to reach with current searches.  Combining this with the precision of atom interferometry, which uses unscreened atoms as the test particles moving in the chameleon field, we have shown that such a configuration allows us to reach previously unobtainable parts of the chameleon parameter space.

\end{document}